# Implementation of The Double-Centroid Reduced Representation of Proteins and its Application to the Prediction of Ligand Binding Sites and Protein-Protein Interaction Partners Using FORTRAN 77/90 Language

**Vicente M. Reyes, Ph.D.***

**E-mail: vmrsbi.RIT.biology@gmail.com**

*work done at:

Dept. of Pharmacology, School of Medicine,
University of California, San Diego
9500 Gilman Drive, La Jolla, CA 92093-0636

&

Dept. of Biological Sciences, School of Life Sciences
Rochester Institute of Technology
One Lomb Memorial Drive, Rochester, NY 14623

**Abbreviations:** AAR, all-atom representation; DCRR, double-centroid reduced represernttion; 3D SM, three-dimensional search motif; ISMTP, interface search motif tetrahedral pair; LBS, ligand binding site; HTP, high-throughput; PLI, protein-ligand interaction; PPI, protein-protein interaction; PDB, Protein Data Bank

## 1. ABSTRACT:

Transformation of protein 3D structures from their all-atom representation (AAR) to the double-centroid reduced representation (DCRR) is a prerequisite to the implementation of both the tetrahedral three-dimensional search motif (3D SM) method for detecting/predicting specific ligand binding sites (LBS) in proteins, and the 3D interface search motif tetrahedral pair (3D ISMTP) method for determining potential binary protein-protein interaction (PPI) partners (Reyes, V.M., 2015a, 2015b, 2009a and 2009b; Reyes, V.M., 2015c and Reyes, V.M., 2009c). In this report we describe results demonstrating the efficacy of the set of FORTRAN 77 and 90 source codes used in the transformation from AAR to DCRR and the implementation of the 3D SM and 3D ISMTP methods. Precisely, we show here the construction of the 3D SM for the biologically important ligands, GTP (of the small Ras-type G-protein family) and sialic acid, from a training set composed of experimentally solved structures of proteins complexed with the pertinent ligand, and their subsequent use in the screening for potential receptor proteins of the two ligands. We also show here the construction of the 3D ISMTP for the binary complexes, RAC:P67PHOX and KAP:phospho-CDK2, from a training set composed of the experimentally solved complexes, and their subsequent use in the screening for potential protomers of the two complexes. The 15 FORTRAN program source codes used in the AAR to DCRR transformation and the implementation of the two aforementioned methods, all presented here in text format, are: (1.) get_bbn.f, (2.) get_sdc.f, (3.) res2cm_bbn.f, (4.) res2cm_sdc.f, (5.) nrst_ngbr.f, (6.) find_Hbonds.f, (7.) find_VDWints.f, (8.) find_clusters.f90, (9.) find_trees.f90, (10.) find_edgenodes.f90, (11.) match_nodes.f, (12.) fpBS.f90, (13.) Gen_Chain_Separ.f, (14.) remove_H_atoms.f and (15.) resd_num_reduct.f. A couple of flowcharts (of programs, inputs and outputs) - one showing how to implement the tetrahedral 3D SM method to find LBSs in proteins, and another showing how to implement the 3D ISMTP method to find binary PPI partners - are presented in our two companion papers (Figure 2 of Reyes, V.M., 2015a, and Figures 1 and 2 of Reyes, V.M., 2015c).

## 2. INTRODUCTION:

The prediction of a protein's function from its 3D, or tertiary, structure is quite an important capability in this day and age of high-throughput (HTP) protein 3D structure determination: most novel proteins get their 3D structures determined before they reach the wet laboratory bench for the determination of their biological functions (please see Introduction sections of Reyes, V.M., 2015a, 2015b and 2015c, and the review references cited therein). There are a least two approaches to the prediction of a protein's function based (entirely) on its 3D structure. One is to determine which ligand(s) the protein binds, and the second is to determine which other protein(s) the protein in question interacts with (*Ibid.*). The first is a question of protein-ligand interactions (PLI), while the second is a question of protein-protein interactions (PPI). The program source codes presented here may be applied to both PLI (Reyes, V.M., 2015a and 2015b) and PPI (Reyes, V. M., 2015c) through the use of the protein DCR representation (Reyes, V.M. & Sheth, V.N., 2011 & 2013).

Source program codes presented in this work were written in either Fortran 77 (Holoien, M.O. & Behforooz, A., 1991; Mayo, W. & Cwiakala, M., 1994; and Nyhoff, L. & Leestma, S., 1996) or Fortran 90 ( Nyhoff, L. & Leestma, S., 1996 & 1999; Metcalf, M. & Reid, J.K., 1999; and Chapman, S.J., 1997). All program runs were carried out on a UNIX computing environment with a Fortran compiler software. In order to apply the procedure in HTP batch mode, UNIX C-shell (Powers, S. et al., 2002; Anderson, G. & Anderson, P., 1986; and Birns, P. et al., 1985) as well as Perl (Tisdall, J., 2001; and Berman, J.J., 2007) scripts were written. In some complex cases, the scripts were constructed using text manipulation by sed & awk (Dougherty, D. & Robbins, A., 1997; and Aho et al., 1988).

## 3. DATASETS AND METHODS:

The Protein Data Bank (PDB) is the main dataset upon which the programs presented in this paper were applied. The PDB is the main international repository for protein 3D structures (Berman et al., 2000). For specific datasets, please refer to the main references cited in the previous and foregoing sections, particularly Reyes,

V.M. (2015a, 2015b and 2015c). It is strongly suggested that this paper be read in conjunction with Reyes, V.M. (2015a) in order for the reader to see precisely how the programs presented here are applied.

The minimum requirements in running these program source codes is a UNIX computing environment and a Fortran 77/90 compiler software. All Fortran program source codes are compiled before they are run. Application of the procedure to a large dataset of protein structures will be require a knowledge of scripting methods such as UNIX C-shell scripting and Perl programming, as well as that of sed and awk for text manipulation.

## 4. RESULTS AND DISCUSSION:

The Fortran program source codes presented in this paper are tabulated below; page numbers refer to the present paper. Program names ending in suffix ".f" are Fortran 77 codes, while those ending in ".f90" are Fortran 90 codes. We refer the reader to our previous publication, namely, Reyes, V.M., 2015a, 2015b and 2015c, for the implementation and application of these programs in an actual research setting, specifically, what each program accomplishes. Please refer to Figures 2 and 5B of Reyes, V.M. (2015a) for a flowchart and illustration of how these programs are implemented.

**Table of Programs**

### 4.1 Implementation in the Prediction of Protein-Ligand Interactions.

Use of DCRR to represent proteins dramatically reduces the atomicity of the protein structure. As an illustration, compare the atomic structure of the octapeptide OM99-2 (glu-val-asn-leu-ala-ala-glu-phe; a memepsin inhibitor) on the upper panel of Figure 1 with that of its DCRR representation on the lower panel. The atomicities of the eight amino acids in the all-atom representation ranges from five (ala) to 11 (phe) while in the DCRR, their atomicities have all been drastically reduced to two. Table 1 shows the structure file for a leucine residue in AAR on the upper panel, and compares it to one transformed to DCRR on the lower panel. Note that each residue is represented by two coordinates, represented by the square - the coordinate of the centroid of the backbone atoms, N, CA, C' and O; and by the triangle – the coordinates of the centroid of the sidechain atoms, CB, CG, CD, etc.

Figure 2 illustrates a ligand bound at its binding site in the receptor protein in AAR on the top panel, and on the lower panel, it shows the same scenario but with the protein receptor in DCRR. This reduction of atomicity and thus complexity in the ligand/LBS structure allowed us to model the interaction mathematically. This model is embodied in the data structure called the three-dimensional search motif (3D SM) shown in Figure 3. This model is composed of four points in 3D space, and is this generally a tetrahedron. This model is information-rich and in general contains eight qualitative and six quantitative parameters for a total of 14 parameters. The eight qualitative parameters are due to the amino acid identities of the four vertices of the tetrahedron and whether each is a backbone or sidechain centroid. The six quantitative parameters are the lengths of the six sides (edges) of the tetrahedron. This abundance of information content in the search motif gives our procedure high specificity.

A prerequisite to being able to model the PLI at the LBS is to determine the interactions between ligand atoms and protein atoms, i.e., the hydrogen bonds and van der Waals interactions in the ligand-bound structure. This is shown schematically in Figure 4. We obtain these information from 'training structures'. Training structures are experimentally solved structures which are analyzed to extract the screening parameters from. We perform a nearest neighbor analysis with the ligand as the "home" file and the protein as the "neighbor" file, from which we obtain the nearest neighbors (in the protein) of the ligand atoms. From these nearest neighbors, we select the H-bonds and VDW interactions using programs find_Hbonds.f and find_VDWints.f. The four most dominant ones are then chosen as the nodes of the tetrahedron. Converting the protein to DCRR finally gives us the 3D SM. Although it probably is not critical, we choose the root vertex R to be the most dominant of the four, and the other three are nodes 1, 2 and 3.

We now apply our procedure to real protein-ligand interactions. The 16 training structures we used for GTP binding site screening in the srGP family is shown in Table 2, while those used for SIA are shown in Table 3. Note that all of them have the appropriate ligand (GTP and sialic acid, respectively) bound in them. In Figure 6 we show the 3D SM and its parameters for the srGP and sialic acid (SIA). For purposes of specification in the programs, we designate the root-node sides, e.g., Rn1, Rn2 and Rn3, as "branches", and the node-node sides as "node-edges".

Our test set is a set of 801 protein structures in the PDB whose functions were undetermined (as of 2006; Reyes, V.M., 2015a, 2015b). The final screening results and predictions for both srGP and SIA are shown in Table 5. Note that two, namely, 1RUB and 1XTL, of the 801 test structures were predicted to be a member of the srGP family, while four, namely, 1IUK, 1SQH, 1VKA and 1Y6Z, were predicted to bind SIA. In particular, note that 1VKA is of human origin, and may be a novel drug target.

### 4.2 Implementation in the Prediction of Protein-Protein Interaction Partners.

Our ligand BS screening procedure may be extended to protein-protein interactions (PPI) to predict PPI partners. Consider the interactions at the interface of a binary protein complex made up of protomers A and B. As shown in Figure 5, at the interface of these two proteins are two tetrahedra interacting via H-bonds and van der Waals attractive forces. If we consider each of these two tetrahedral as a 3D SM (as in the previous section) on PLI, then by screening one's test set twice – once for each interfacial 3D SM – then we can predict PPI partners using the same procedure as in predicting LBS's in PLI.

We applied this procedure to real proteins, namely, our test set of 801 protein structures in the PDB with unknown functions as mentioned above (Reyes, V.M., 2015a, 2015b). The training structures in this case must be binary complexes whose structures have been solved experimentally. The training structures for complexes D and H (each composed of protomers 1 and 2) are shown in Table 4. Note that there are only one training structure for each complex; this should suffice. Figure 7 shows the 3D ISMTP and their parameters for these two complexes. In the diagram, the solid blue and red lines represent intra-protomer interactions in protomers #1 and #2 in the two complexes, while the green broken lines are represent inter-protomer interactions in the complexes. Note that the ISMTP has a total of 32 parameters: 14 each for the two tetrahedra (2x14=28) and 4 inter-protomer lengths, R-R', n1-n1', n2-n2' and n3-n3' (28+4 = 32).

Our screening results reveal the following positive results: For complex D, there were eight positive structures for protomer #1 and six positive structures for protomer #2. That makes a total of 8x6 = 48 possible candidate complexes D from the test set. As for complex H, there are eight positive structures for protomer #1 and 15 structures positive for protomer #2. This makes a total of 8x15 = 120 possible complexes H from the test set. These positive structures each represent roughly 1.0% - 2.0% of the test set (which had 801 elements).

The final screening results and predictions for complexes D and H are shown in Table 6. The two candidates for protomer #1 of complex D are 1J2R and 2F4L; while the four candidates for protomer #2 are 1SBK, 1VHS, 1VI8 and 1ZBR. The lone candidates for protomer #1 of complex H is 1XVS; while the three candidates for protomer #2 are 1S7O, 1V99 and 1XG8.

### 4.2 Conclusion.

In conclusion, we have shown that the 15 FORTRAN program source codes presented in this paper do perform their expected overall functions effectively. We also conclude that the 3D SM and 3D ISMTP methods for predicting protein-ligand and protein-protein interactions using the programs presented here are valid and robust. Our objective for the foreseeable future is to apply the procedures described in this and the companion papers (Reyes, V.M., 2015a, 2015b) in large-scale, especially to all proteins whose structures have been solved and deposited in the PDB but whose biological functions are still unknown.

## 5. ACKNOWLEDGMENT:

This work was supported by an Institutional Research and Academic Career Development Award to the author, NIGMS/NIH grant number GM 68524. The author wishes to acknowledge the San Diego Supercomputer Center, the UCSD Academic Computing Services, and the UCSD Biomedical Library, for the help and support of their staff and personnel. He also acknowledges the Division of Research Computing at RIT, and computing resources from the Dept. of Biological Sciences, College of Science, at RIT.

**7. FIGURES and LEGENDS:**

---

---

## 8. TABLES and LEGENDS:

---

---

## 9. PROGRAMS:

**Program 1:**

**##########    Start of Pogram   "get_bbn.f"   ##########**

```
c     program name: get_bbn.f

c c c c c c c c c c c c c c c c c c c c c c c c c c c c c c c c c c c c c
c                                                                       c
c   Author:  Vicente M. Reyes, Ph.D.                                    c
c            Dept. of Pharmacol., Skaggs Sch. of Pharm. & Pharm. Sci.   c
c         &  Dept. of Integrative Biosci., S.D. Supercomptr. Ctr.       c
c            La Jolla, CA  92093-0505  U.S.A.                           c
c                                                                       c
c c c c c c c c c c c c c c c c c c c c c c c c c c c c c c c c c c c c c

      character*4 labla
      character*8 linna
      character*5 atoma
      character*65 right_side

c     character*3 resda
c     character*1 chaina
c     integer linna
c     integer resnoa
c     real xa, ya , za, occa, bfa


      open (unit =1, file = "filei")
      open (unit =2, file = "fileo")

888   read(1,100,end=333) labla, linna, atoma, right_side
c                                           resda,chaina,resnoa,xa,ya,za,occa,bfa

      if ((atoma.eq.'N   ').or.
     +(atoma.eq.'CA  ').or.
     +(atoma.eq.'C   ').or.
     +(atoma.eq.'O   ')) then

      write(2,100) labla, linna, atoma, right_side
c                                     resda,chaina,resnoa,xa,ya,za,occa,bfa
```

```
      endif

      go to 888

100   format(A4,A8,A5,A65)
c                                A3,1x,A1,I5,3x,f8.3,f8.3,f8.3,2x,f4.2,f6.2)

333   continue

      close(2)
      close(1)

      stop
      end


c     character*30 misc, misca, miscb
c     character*8 resno, resnoa, resnob
c     character*4 labl, labla, lablb
c     character*8 linn, linna, linnb
c     character*5 atom, atoma, atomb
c     character*4 resd, resda, resdb
c     character*1 chn, chna, chnb
c     real count,x,y,z,sumx,sumy,sumz,avex,avey,avez
```

**##########   End of Pogram   "get_bbn.f"   ##########**

**Program 2:**

**##########   Start of Pogram   "get_sdc.f"  ##########**

```
c     program name: get_sdc.f

c c c c c c c c c c c c c c c c c c c c c c c c c c c c c c c c c c c c c c
c                                                                        c
c   Author:  Vicente M. Reyes, Ph.D.                                     c
c            Dept. of Pharmacol., Skaggs Sch. of Pharm. & Pharm. Sci.    c
c         &  Dept. of Integrative Biosci., S.D. Supercomptr. Ctr.        c
c            La Jolla, CA  92093-0505  U.S.A.                            c
c                                                                        c
c c c c c c c c c c c c c c c c c c c c c c c c c c c c c c c c c c c c c c

      character*4 labla
      character*8 linna
      character*5 atoma
      character*65 right_side

c     character*3 resda
c     character*1 chaina
c     integer linna
c     integer resnoa
c     real xa, ya , za, occa, bfa


      open (unit =1, file = "filei")
      open (unit =2, file = "fileo")

888   read(1,100,end=333) labla, linna, atoma, right_side
c
resda,chaina,resnoa,xa,ya,za,occa,bfa

c      if ((atoma.eq.'N   ').or.
c    +(atoma.eq.'CA  ').or.
```

```
c    +(atoma.eq.'C   ').or.
c    +(atoma.eq.'O   ')) then

      if (.not.((atoma.eq.'N   ').or.
     +(atoma.eq.'CA  ').or.
     +(atoma.eq.'C   ').or.
     +(atoma.eq.'O   '))) then


      write(2,100) labla, linna, atoma, right_side
c                                  resda,chaina,resnoa,xa,ya,za,occa,bfa

      endif

      go to 888

100   format(A4,A8,A5,A65)
c                            A3,1x,A1,I5,3x,f8.3,f8.3,f8.3,2x,f4.2,f6.2)

333   continue

      close(2)
      close(1)

      stop
      end

c     character*30 misc, misca, miscb
c     character*8 resno, resnoa, resnob
c     character*4 labl, labla, lablb
c     character*8 linn, linna, linnb
c     character*5 atom, atoma, atomb
c     character*4 resd, resda, resdb
c     character*1 chn, chna, chnb
c     real count,x,y,z,sumx,sumy,sumz,avex,avey,avez
```

**##########   End of Pogram   "get_sdc.f" ##########**

**Program 3:**

**##########   Start of Pogram  "res2cm_bbn.f"   ##########**

```
c  program name: res2cm_bbn.f

c c c c c c c c c c c c c c c c c c c c c c c c c c c c c c c c c c c c c c
c                                                                        c
c   Author:  Vicente M. Reyes, Ph.D.                                     c
c            Dept. of Pharmacol., Skaggs Sch. of Pharm. & Pharm. Sci.    c
c         &  Dept. of Integrative Biosci., S.D. Supercomptr. Ctr.        c
c            La Jolla, CA  92093-0505  U.S.A.                            c
c                                                                        c
c c c c c c c c c c c c c c c c c c c c c c c c c c c c c c c c c c c c c c

c  from a pdb file, this program calculates the center of mass of each residue
c  treating each atom in the residue to be of the same mass, then replaces the
c  entire residue with the center of mass
c  Pre-processing required: duplicate the **last** line of the input pdb file,
c  then relabel its residue number "0"; this is the dummy residue with residue
c  number "0"

      character*30 misc, misca, miscb
      character*8 resno, resnoa, resnob
      character*4 labl, labla, lablb
```

```
      character*8 linn, linna, linnb
      character*5 atom, atoma, atomb
      character*4 resd, resda, resdb
      character*1 chn, chna, chnb
      real count,x,y,z,sumx,sumy,sumz,avex,avey,avez

      open (unit =1, file = "filei")
      open (unit =2, file = "fileo")

888   read(1,100,end=333) labl,linn,atom,resd,
     +  chn,resno,x,y,z,misc

       labla = labl
       linna = linn
       atoma = atom
       resda = resd
       chna = chn
       resnoa = resno
       misca = misc

       sumx = x
       sumy = y
       sumz = z

       count = 1.0

999   read(1,100,end=333) labl,linn,atom,resd,
     +   chn,resno,x,y,z,misc

       lablb = labl
       linnb = linn
       atomb = atom
       resdb = resd
       chnb = chn
       resnob = resno
       miscb = misc

      if ((resda.eq.resdb).and.(resnoa.eq.resnob)
     +     .and.(chna.eq.chnb)) then

      sumx = sumx + x
      sumy = sumy + y
      sumz = sumz + z

      count = count + 1.0

       labla = lablb
       linna = linnb
       atoma = atomb
       resda = resdb
       chna = chnb
       resnoa = resnob
       misca = miscb

      go to 999

      else

      avex = sumx/count
      avey = sumy/count
      avez = sumz/count

      write (2,100) labla, linna, ' bbc ', resda,
     +   chna,  resnoa,avex, avey, avez, misca

       labla = lablb
```

```
        linna = linnb
        atoma = atomb
        resda = resdb
        chna = chnb
        resnoa = resnob
        misca = miscb

      count = 1.0
      sumx = x
      sumy = y
      sumz = z

      go to 999

      endif

c     character*30 misc, misca, miscb
c     character*8 resno, resnoa, resnob
c     character*4 labl, labla, lablb
c     character*8 linn, linna, linnb
c     character*5 atom, atoma, atomb
c     character*4 resd, resda, resdb
c     character*1 chn, chna, chnb
c     real count,x,y,z,sumx,sumy,sumz,avex,avey,avez

100   format(A4, A8, A5, A4, A1, A8, f8.3, f8.3, f8.3, A30)

333   continue

      close(2)
      close(1)

      stop
      end
```

########## End of Pogram "res2cm_bbn.f" ##########

**Program 4:**

########## Start of Pogram "res2cm_sdc.f" ##########

```
c  program name: res2cm_sdc.f

c c c c c c c c c c c c c c c c c c c c c c c c c c c c c c c c c c c c c c
c                                                                         c
c   Author:  Vicente M. Reyes, Ph.D.                                      c
c            Dept. of Pharmacol., Skaggs Sch. of Pharm. & Pharm. Sci.     c
c         &  Dept. of Integrative Biosci., S.D. Supercomptr. Ctr.         c
c            La Jolla, CA  92093-0505  U.S.A.                             c
c                                                                         c
c c c c c c c c c c c c c c c c c c c c c c c c c c c c c c c c c c c c c c

c  from a pdb file, this program calculates the center of mass of each residue
c  treating each atom in the residue to be of the same mass, then replaces the
c  entire residue with the center of mass
c  Pre-processing required: duplicate the **last** line of the input pdb file,
c  then relabel its residue number "0"; this is the dummy residue with residue
c  number "0"

      character*30 misc, misca, miscb
      character*8 resno, resnoa, resnob
```

```
      character*4 labl, labla, lablb
      character*8 linn, linna, linnb
      character*5 atom, atoma, atomb
      character*4 resd, resda, resdb
      character*1 chn, chna, chnb
      real count,x,y,z,sumx,sumy,sumz,avex,avey,avez

      open (unit =1, file = "filei")
      open (unit =2, file = "fileo")

888   read(1,100,end=333) labl,linn,atom,resd,
     +  chn,resno,x,y,z,misc

       labla = labl
       linna = linn
       atoma = atom
       resda = resd
       chna = chn
       resnoa = resno
       misca = misc

       sumx = x
       sumy = y
       sumz = z

       count = 1.0

999   read(1,100,end=333) labl,linn,atom,resd,
     +   chn,resno,x,y,z,misc

       lablb = labl
       linnb = linn
       atomb = atom
       resdb = resd
       chnb = chn
       resnob = resno
       miscb = misc

      if ((resda.eq.resdb).and.(resnoa.eq.resnob)
     +     .and.(chna.eq.chnb)) then

      sumx = sumx + x
      sumy = sumy + y
      sumz = sumz + z

      count = count + 1.0

       labla = lablb
       linna = linnb
       atoma = atomb
       resda = resdb
       chna = chnb
       resnoa = resnob
       misca = miscb

      go to 999

      else

      avex = sumx/count
      avey = sumy/count
      avez = sumz/count

      write (2,100) labla, linna, ' sdc ', resda,
     +   chna,  resnoa,avex, avey, avez, misca
```

```
        labla = lablb
        linna = linnb
        atoma = atomb
        resda = resdb
        chna = chnb
        resnoa = resnob
        misca = miscb

      count = 1.0
      sumx = x
      sumy = y
      sumz = z

      go to 999

      endif

c     character*30 misc, misca, miscb
c     character*8 resno, resnoa, resnob
c     character*4 labl, labla, lablb
c     character*8 linn, linna, linnb
c     character*5 atom, atoma, atomb
c     character*4 resd, resda, resdb
c     character*1 chn, chna, chnb
c     real count,x,y,z,sumx,sumy,sumz,avex,avey,avez

100   format(A4, A8, A5, A4, A1, A8, f8.3, f8.3, f8.3, A30)

333   continue

      close(2)
      close(1)

      stop
      end
```

**##########  End of Pogram  "res2cm_sdc.f" ##########**

**Program 5:**

**##########  Start of Pogram "nrst_ngbr.f" ##########**

```
c  program name: nrst_ngbr.f

c c c c c c c c c c c c c c c c c c c c c c c c c c c c c c c c c c c c c
c                                                                       c
c   Author:  Vicente M. Reyes, Ph.D.                                    c
c            Dept. of Pharmacol., Skaggs Sch. of Pharm. & Pharm. Sci.   c
c          & Dept. of Integrative Biosci., S.D. Supercomptr. Ctr.       c
c            La Jolla, CA  92093-0505  U.S.A.                           c
c                                                                       c
c c c c c c c c c c c c c c c c c c c c c c c c c c c c c c c c c c c c c

c  filea is the "home" pdb file;
c  fileb is the "neighbor" pdb file;
c  fileo is the results file


      character*10 misca, miscb
      character*6 labla, lablb
      character*5 linna, linnb, resnoa, resnob
```

```
      character*4 atoma, atomb
      character*3 resda, resdb
      character*1 chaina, chainb
      real xa, xb, ya , yb, za, zb, dist, lim

      lim = 5.000

      open (unit =1, file = "filea")
      open (unit =2, file = "fileb")
      open (unit =3, file = "fileo")

888   read(1,100,end=333) labla, linna, atoma, resda, chaina, resnoa,
     +                    xa, ya, za, misca
999   read(2,100,end=444) lablb, linnb, atomb, resdb, chainb, resnob,
     +                    xb, yb, zb, miscb

      dist = sqrt((xa-xb)**2 + (ya-yb)**2 + (za-zb)**2)

      if(dist.le.lim) then
      write(3,200) atoma,resda,chaina,resnoa, atomb,resdb,chainb,
     +                                  resnob, dist
      endif

      go to 999

444   rewind 2

      go to 888

100   format(A6,A5,2x,A4,A3,x,A1,A5,3x,
     +     f8.3,f8.3,f8.3,2x,A10)

200   format(A4,x,A3,1x,A1,A5,2x,A4,x,A3,1x,A1,A5,2x,f7.3)

333   continue

      close(3)
      close(2)
      close(1)

      stop
      end
```

**##########    End of Pogram  "nrst_ngbr.f"  ##########**

**Program 6:**

**##########    Start of Pogram  "find_HBonds.f"   ##########**

```
c     program name: find_Hbonds.f

c c c c c c c c c c c c c c c c c c c c c c c c c c c c c c c c c c c c c c
c                                                                         c
c   Author:  Vicente M. Reyes, Ph.D.                                      c
c            Dept. of Pharmacol., Skaggs Sch. of Pharm. & Pharm. Sci.     c
c         &  Dept. of Integrative Biosci., S.D. Supercomptr. Ctr.         c
c            La Jolla, CA  92093-0505  U.S.A.                             c
c                                                                         c
c c c c c c c c c c c c c c c c c c c c c c c c c c c c c c c c c c c c c c

c     input to this program is the output of the nrst_ngbr.f program
```

```
      character*14 A_segment, B_segment
      character*1 A_atom, B_atom
      real dist

      open (unit =1, file = "filei")
      open (unit =2, file = "fileo")

888   read(1,100,end=333) A_atom,A_segment,B_atom,B_segment,dist

      if (((dist.le.3.2).and.(dist.ge.2.5)).and.((A_atom.eq.'N').or.
     + (A_atom.eq.'O').or.(A_atom.eq.'S')).and.((B_atom.eq.'N').or.
     + (B_atom.eq.'O').or.(B_atom.eq.'S'))) then

      write(2,100) A_atom,A_segment,B_atom,B_segment,dist

      endif

      go to 888

100   format(A1, A14, 2x, A1, A14, 2x, f7.3)

333   continue

      close(2)
      close(1)

      stop
      end
```

**########## End of Pogram "find_HBonds.f" ##########**

**Program 7:**

**########## Start of Pogram "find_VDWints.f" ##########**

```
c     program name: find_VDWints.f

c c c c c c c c c c c c c c c c c c c c c c c c c c c c c c c c c c c c c c
c                                                                         c
c   Author:  Vicente M. Reyes, Ph.D.                                      c
c            Dept. of Pharmacol., Skaggs Sch. of Pharm. & Pharm. Sci.    c
c         &  Dept. of Integrative Biosci., S.D. Supercomptr. Ctr.        c
c            La Jolla, CA  92093-0505  U.S.A.                             c
c                                                                         c
c c c c c c c c c c c c c c c c c c c c c c c c c c c c c c c c c c c c c c

      character*14 A_segment, B_segment
      character*1 A_atom, B_atom
      real dist

      open (unit =1, file = "filei")

      open (unit =2, file = "filev")
      open (unit =3, file = "filew")
      open (unit =4, file = "filex")
      open (unit =5, file = "filey")
      open (unit =6, file = "filez")


888   read(1,100,end=333)A_atom, A_segment, B_atom, B_segment, dist
```

```
100   format(A1, A14, 2x, A1, A14, 2x,  f7.3)

      if (((dist.ge.3.30).and.(dist.le.4.80)).and.
     +   ((A_atom.eq.'C') .and.(B_atom.eq.'C'))) then

         write (2,100)A_atom, A_segment, B_atom, B_segment, dist


      elseif (((dist.ge.3.12).and.(dist.le.4.62)).and.
     +    (((A_atom.eq.'C').and.(B_atom.eq.'O')).or.
     +    ((A_atom.eq.'O').and.(B_atom.eq.'C')))) then

        write (3,100)A_atom, A_segment, B_atom, B_segment, dist


      else if (((dist.ge.3.15).and.(dist.le.4.65)).and.
     +  (((A_atom.eq.'C').and.(B_atom.eq.'N')).or.
     +  ((A_atom.eq.'N').and.(B_atom.eq.'C')))) then

        write (4,100)A_atom, A_segment, B_atom, B_segment, dist


      elseif (((dist.ge.3.40).and.(dist.le.4.90)).and.
     + (((A_atom.eq.'C').and.  (B_atom.eq.'S')).or.
     + ((A_atom.eq.'S').and.(B_atom.eq.'C')))) then

        write  (5,100)A_atom, A_segment, B_atom, B_segment, dist


      elseif (((dist.ge.3.40).and.(dist.le.4.90)).and.
     + (((A_atom.eq.'C').and.  (B_atom.eq.'P')).or.
     + ((A_atom.eq.'P').and.(B_atom.eq.'C')))) then

         write(6,100)A_atom, A_segment, B_atom, B_segment, dist

      endif

      go to 888

333   continue

      close(6)
      close(5)
      close(4)
      close(3)
      close(2)
      close(1)

      stop
      end
```

**########## End of Pogram "find_VDWints.f" ##########**

**Program 8:**

**########## Start of Pogram "find_clusters.f90" ##########**

```
program find_clusters

implicit none

character(12), dimension(0:124):: left
character(30), dimension(0:124):: right
character(2), dimension(0:124):: tag
integer, dimension(0:124):: resno
integer:: i, j, k, l

open (unit=10, file="filei")
open (unit=12, file="fileo")

100 format(A12,I4,A30,A2)

Singlet: do i = 0,124,1
    read(unit=10,fmt=100) left(i),resno(i),right(i),tag(i)
end do  Singlet

!   print*, left(0), resno(8), right(28), tag(58)

j = 0
88  k = 0

if (j >=124) then
go to 77
else
go to 99
end if

99  if (resno(j) == resno(j+1)) then
        j = j + 1
        k = k + 1

!       print*, resno(j)

        go to 99

    else

        if (k >= 2) then
            Inner: do l = j-k,j,1
                write (12,100) left(l),resno(l),right(l),tag(l)
            end do Inner
        else
            j = j + 1
            go to 88
        end if

        go to 88

    end if

77    close(10)
      close(12)

end program find_clusters
```

**##########    End of Pogram   "find_clusters.f90"   ##########**

**Program 9:**

########## Start of Pogram "find_trees.f90" ##########

```
program find_trees

implicit none

character(12), dimension(0:88):: left
character(30), dimension(0:88):: right
character(2), dimension(0:88):: tag
integer, dimension(0:88):: resno
integer:: a, b, c, i, j, k, m, n

open (unit=10, file="filei")
open (unit=12, file="fileo")

100 format(A12,I4,A30,A2)
101 format(A12,I4,A30,A2,x,I2)

do i = 0,88,1
    read(unit=10,fmt=100) left(i),resno(i),right(i),tag(i)
end do

! print*, left(0), resno(8), right(88), tag(88), tag(89)

n = 0
88  a = 0
    b = 0
    c = 0

if (n >=88) then
go to 77
else
go to 99
end if

99  if (resno(n) == resno(n+1)) then

          if (tag(n).eq.'ba') then
             a = a + 1
          else if (tag(n).eq.'bd') then
             b = b + 1
          else if (tag(n).eq.'bc') then
             c = c + 1
          endif

        n = n + 1
        go to 99

    else

         if (tag(n).eq.'ba') then
             a = a + 1
         else if (tag(n).eq.'bd') then
             b = b + 1
         else if (tag(n).eq.'bc') then
             c = c + 1
         endif

        if ((a.ge.1).and.(b.ge.1).and.(c.ge.1)) then
            do m = n-(a+b+c-1),n,1
                if (tag(m).eq.'ba') then
                write (12,101) left(m),resno(m),right(m),tag(m),a
                else if (tag(m).eq.'bd') then
```

```
               write (12,101) left(m),resno(m),right(m),tag(m),b
               else if (tag(m).eq.'bc') then
               write (12,101) left(m),resno(m),right(m),tag(m),c
               end if
           end do
        else
           n = n + 1
           go to 88
        end if

        go to 88

    end if

77    close(10)
      close(12)

end program find_trees
```

########## End of Pogram "find_tresss.f90" ##########

**Program 10:**

########## Start of Pogram "find_edgenodes.f90" ##########

```
c     program name: find_edgenode.f

c c c c c c c c c c c c c c c c c c c c c c c c c c c c c c c c c c c c c c
c                                                                       c
c   Author:  Vicente M. Reyes, Ph.D.                                   c
c            Dept. of Pharmacol., Skaggs Sch. of Pharm. & Pharm. Sci.   c
c          & Dept. of Integrative Biosci., S.D. Supercomptr. Ctr.      c
c            La Jolla, CA  92093-0505  U.S.A.                          c
c                                                                       c
c c c c c c c c c c c c c c c c c c c c c c c c c c c c c c c c c c c c c c

      character*18 node1, node2
      character*14 rootres1, rootres2
      character*3 tag1, tag2, labl
      character*4 linno1, linno2, linnoR1, linnoR2
      integer rootno1, rootno2
      real dist1, dist2

      labl = 'R ='

c     2 input files:
      open (unit =1, file = "filea")
      open (unit =2, file = "fileb")

c     1 output file:
      open (unit =3, file = "fileo")

888   read(1,100,end=333) rootres1, rootno1, node1, dist1,
     +                    linnoR1, linno1, tag1
999   read(2,100,end=444) rootres2, rootno2, node2, dist2,
     +                    linnoR2, linno2, tag2

100   format(A14,I4,3x,A18,3x,f7.3,x,A4,x,A4,x,A3)

csCM   ARG  -   364   sCM   TYR  -  367      6.179 2826 2851 Rn1  2
```

```
csCM   ARG  -   364   sCM   TYR  -  367      6.269 6043 6068 Rn1  2
csCM   ARG  -   364   sCM   MET  -   62      9.481 2826  470 Rn2  2
csCM   ARG  -   364   sCM   MET  -   62      9.497 6043 3687 Rn2  2
csCM   ARG  -   209   bCM   GLY  -   371     6.304 1592 6094 Rn3  4
csCM   ARG  -   209   bCM   GLY  -   371     6.318 4809 2877 Rn3  4

      if (rootno1.eq.rootno2) then
      write(3,200) node1, node2, labl, rootno1, linnoR1, linnoR2,
     +                    linno1, linno2, tag1, tag2
200   format(A18,x,A18,x,A3,I4,x,A4,x,A4,x,A4,x,A4,x,A3,x,A3)
      endif

      go to 999

444   rewind 2

      go to 888

333   continue

      close(3)
      close(2)
      close(1)

      stop
      end
```

**##########    End of Pogram    "find_edgenodes.f90"    ##########**

**Program 11:**

**###########   Start of Pogram   "match_nodes.f"   #############**

```
c     program name: match_nodes.f

c c c c c c c c c c c c c c c c c c c c c c c c c c c c c c c c c c c c c
c                                                                       c
c   Author:  Vicente M. Reyes, Ph.D.                                    c
c            Dept. of Pharmacol., Skaggs Sch. of Pharm. & Pharm. Sci.   c
c         &  Dept. of Integrative Biosci., S.D. Supercomptr. Ctr.       c
c            La Jolla, CA  92093-0505  U.S.A.                           c
c                                                                       c
c c c c c c c c c c c c c c c c c c c c c c c c c c c c c c c c c c c c c

      character*18 node1a,node1b,node2a,node2b
      character*10 desc1
      character*35 desc2
      real dist1

      open (unit =1, file = "filea")
      open (unit =2, file = "fileb")
      open (unit =3, file = "fileo")

c     unit1 ("filea") is the all.XX
888   read(1,101,end=333) node1a, node1b, dist1, desc1
101   format(A18,3x,A18,3x,f7.3,A10)

c*******************************************************************
csCM   SER  -  104    sCM   MET  -  106      6.368  789  809
csCM   SER  -  104    sCM   MET  -    1      4.314 1574  835
csCM   SER  -  104    sCM   MET  -    1      4.218 2319 1620
```

```
csCM   SER  -  104    sCM   MET  -  106      6.020 2319 2339
c*****************************************************************

c     unit2 ("fileb") is the edgenode_XX
999   read(2,102,end=444) node2a, node2b, desc2
102   format(A18,x,A18,x,A35)

c*****************************************************************
csCM   TYR  -  103  sCM   MET  -   38  R =  26 1830 2575 3122 2665 Rn1 Rn2
csCM   TYR  -  103  sCM   MET  -   38  R =  26  234 2575 1568 2665 Rn1 Rn2
csCM   TYR  -  105  sCM   MET  -   38  R =  26 2575 2575 2331 2665 Rn1 Rn2
c*****************************************************************

      if (((node1a.eq.node2a).and.(node1b.eq.node2b)).or.
     +    ((node1a.eq.node2b).and.(node1b.eq.node2a)))  then

      write(3,200) node2a, node2b, dist1, desc1, desc2
200   format(A18,x,A18,2x,f7.3,2x,A10,2x,A35)

      endif

      go to 999

444   rewind 2

      go to 888


333   continue

      close(3)
      close(2)
      close(1)

      stop
```

**################   End of Pogram   "match_nodes.f"  ################**

**Program 12:**

**##########   Start of Pogram  "fpBS.f90"   ##########**

```
program fpBS

! implicit none

character(63), dimension(0:xxxx):: left
character(20), dimension(0:xxxx):: right
character(7), dimension(0:xxxx):: branch
integer, dimension(0:xxxx):: rootno
integer:: a, b, c, i, m, n

dim = xxxx

open (unit=10, file="filei")
open (unit=12, file="fileo")

100 format(A63,I4,x,A20,A7)
```

```
!*********************************************************************************
*************
!sCM   MET  -   38    bCM   GLY  -    37   R =  26 Rn2 Rn3     5.379
!sCM   MET  -   38    bCM   GLY  -    37   R =  26 Rn2 Rn3     5.670
!sCM   TYR  -  103    sCM   MET  -   38    R =  26 Rn1 Rn2     5.227
!*********************************************************************************
*************
!sCM   MET  -   38  bCM   GLY  -    37    5.379    315  307  R =  26 2575 1830 2665
1912 Rn2 Rn3
!sCM   TYR  -  103  sCM   MET  -   38     5.227    783 1114  R =  26  234 2575 1568
2665 Rn1 Rn2
!sCM   TYR  -  103  sCM   MET  -   38     5.639   3122 1920  R =  26  234 2575 1568
2665 Rn1 Rn2
!*********************************************************************************
*************

do i = 0,dim,1
   read(unit=10,fmt=100) left(i),rootno(i),right(i),branch(i)
end do

n = 0
88  a = 0
    b = 0
    c = 0

if (n >=dim) then
go to 77
else
go to 99
end if

99  if (rootno(n) == rootno(n+1)) then

          if (branch(n).eq.'Rn1 Rn2') then
             a = a + 1
          else if (branch(n).eq.'Rn1 Rn3') then
             b = b + 1
          else if (branch(n).eq.'Rn2 Rn3') then
             c = c + 1
          endif

        n = n + 1
        go to 99

    else

         if (branch(n).eq.'Rn1 Rn2') then
             a = a + 1
         else if (branch(n).eq.'Rn1 Rn3') then
             b = b + 1
         else if (branch(n).eq.'Rn2 Rn3') then
             c = c + 1
         endif

        if ((a.ge.1).and.(b.ge.1).and.(c.ge.1)) then
            do m = n-(a+b+c-1),n,1
                if (branch(m).eq.'Rn1 Rn2') then
                write (12,100) left(m),rootno(m),right(m),branch(m)
                else if (branch(m).eq.'Rn1 Rn3') then
                write (12,100) left(m),rootno(m),right(m),branch(m)
                else if (branch(m).eq.'Rn2 Rn3') then
                write (12,100) left(m),rootno(m),right(m),branch(m)
                end if
            end do
        else
           n = n + 1
```

```
            go to 88
         end if

         go to 88

     end if

77    close(10)
      close(12)

end program fpBS
```

########## End of Pogram "fpBS.f90" ##########

**Program 13:**

**################ Start of Pogram "Gen_Chain_Separ.f" ################**

```
c     program name: GeneralChainSeparation.f

c c c c c c c c c c c c c c c c c c c c c c c c c c c c c c c c c c c c c c
c                                                                        c
c   Author:  Vicente M. Reyes, Ph.D.                                     c
c            Dept. of Pharmacol., Skaggs Sch. of Pharm. & Pharm. Sci.    c
c         &  Dept. of Integrative Biosci., S.D. Supercomptr. Ctr.        c
c            La Jolla, CA  92093-0505  U.S.A.                            c
c                                                                        c
c c c c c c c c c c c c c c c c c c c c c c c c c c c c c c c c c c c c c c

      character*1 chnID
      character*21 left
      character*38 right

      open (unit =1, file = "filei")

      open (unit =2, file = "file02")
      open (unit =3, file = "file03")
      open (unit =4, file = "file04")
      open (unit =6, file = "file06")
      open (unit =7, file = "file07")
      open (unit =8, file = "file08")
      open (unit =9, file = "file09")
      open (unit =10, file = "file10")
      open (unit =11, file = "file11")
      open (unit =12, file = "file12")
      open (unit =13, file = "file13")
      open (unit =14, file = "file14")
      open (unit =15, file = "file15")
      open (unit =16, file = "file16")
      open (unit =17, file = "file17")
      open (unit =18, file = "file18")
      open (unit =19, file = "file19")
      open (unit =20, file = "file20")
      open (unit =21, file = "file21")
      open (unit =22, file = "file22")
      open (unit =23, file = "file23")
      open (unit =24, file = "file24")
      open (unit =25, file = "file25")
      open (unit =26, file = "file26")
      open (unit =27, file = "file27")
      open (unit =28, file = "file28")

888   read(1,100,end=333) left, chnID, right
```

```
100   format(A21,A,A38)

        if (chnID.eq.'A') then
      write(2,100) left, chnID, right

        elseif (chnID.eq.'B') then
      write(3,100) left, chnID, right

        elseif (chnID.eq.'C') then
      write(4,100) left, chnID, right

        elseif (chnID.eq.'D') then
      write(6,100) left, chnID, right

        elseif (chnID.eq.'E') then
      write(7,100) left, chnID, right

        elseif (chnID.eq.'F') then
      write(8,100) left, chnID, right

        elseif (chnID.eq.'G') then
      write(9,100) left, chnID, right

        elseif (chnID.eq.'H') then
      write(10,100) left, chnID, right

        elseif (chnID.eq.'I') then
      write(11,100) left, chnID, right

        elseif (chnID.eq.'J') then
      write(12,100) left, chnID, right

        elseif (chnID.eq.'K') then
      write(13,100) left, chnID, right

        elseif (chnID.eq.'L') then
      write(14,100) left, chnID, right

        elseif (chnID.eq.'M') then
      write(15,100) left, chnID, right

        elseif (chnID.eq.'N') then
      write(16,100) left, chnID, right

        elseif (chnID.eq.'O') then
      write(17,100) left, chnID, right

        elseif (chnID.eq.'P') then
      write(18,100) left, chnID, right

        elseif (chnID.eq.'Q') then
      write(19,100) left, chnID, right

        elseif (chnID.eq.'R') then
      write(20,100) left, chnID, right

        elseif (chnID.eq.'S') then
      write(21,100) left, chnID, right

        elseif (chnID.eq.'T') then
      write(22,100) left, chnID, right

        elseif (chnID.eq.'U') then
      write(23,100) left, chnID, right
```

```
        elseif (chnID.eq.'V') then
      write(24,100) left, chnID, right

        elseif (chnID.eq.'W') then
      write(25,100) left, chnID, right

        elseif (chnID.eq.'X') then
      write(26,100) left, chnID, right

        elseif (chnID.eq.'Y') then
      write(27,100) left, chnID, right

        elseif (chnID.eq.'Z') then
      write(28,100) left, chnID, right



      endif

      go to 888

333   continue

      close(28)
      close(27)
      close(26)
      close(25)
      close(24)
      close(23)
      close(22)
      close(21)
      close(20)
      close(19)
      close(18)
      close(17)
      close(16)
      close(15)
      close(14)
      close(13)
      close(12)
      close(11)
      close(10)
      close(9)
      close(8)
      close(7)
      close(6)
      close(4)
      close(3)
      close(2)
      close(1)

      stop

      end
```

**################  End of Pogram  "Gen_Chain_Separ.f.f"  ################**

**Program 14:**

**################  Start of Pogram  "remove_H_atoms.f"  ################**

```
c     program name: remove_Hatoms.f
```

```
      character*12 left
      character*1 atom1
      character*1 atom2
      character*66 right

      open (unit =1, file = "filei")
      open (unit =2, file = "fileo")

888   read(1,100,end=333) left,atom1,atom2,right

      if (.not.((atom1.eq.'H').or.(atom2.eq.'H')))then
      write(2,100) left, atom1, atom2, right
      endif


cccccccccccccccccccccccccccccccccccccccccccccccccccccc
cATOM      5  H1  GLY A   1      -0.536   2.996  -9.200  1.00  0.00           H
cATOM      6  H2  GLY A   1      -1.495   2.693 -10.562  1.00  0.00           H
cATOM      7  H3  GLY A   1      -1.187   4.292 -10.086  1.00  0.00           H
cATOM      8  HA2 GLY A   1      -2.998   4.271  -8.814  1.00  0.00           H
cATOM      9  HA3 GLY A   1      -3.291   2.596  -9.296  1.00  0.00           H
cATOM     10  N   ILE A   2      -3.175   2.781  -6.587  1.00  0.00           N
cATOM     11  CA  ILE A   2      -2.977   2.343  -5.165  1.00  0.00           C
cATOM     12  C   ILE A   2      -1.839   1.346  -4.961  1.00  0.00           C
cATOM     13  O   ILE A   2      -0.981   1.615  -4.153  1.00  0.00           O
cATOM     14  CB  ILE A   2      -4.339   1.743  -4.587  1.00  0.00           C
cATOM     15  CG1 ILE A   2      -5.260   2.982  -4.237  1.00  0.00           C
cATOM     16  CG2 ILE A   2      -4.112   0.876  -3.299  1.00  0.00           C
cATOM     17  CD1 ILE A   2      -6.351   2.670  -3.190  1.00  0.00           C
cATOM     18  H   ILE A   2      -4.069   3.054  -6.871  1.00  0.00           H
cATOM     19  HA  ILE A   2      -2.736   3.231  -4.599  1.00  0.00           H
cATOM     20  HB  ILE A   2      -4.817   1.130  -5.340  1.00  0.00           H
cATOM     21 HG12 ILE A   2      -4.659   3.781  -3.835  1.00  0.00           H
cATOM     22 HG13 ILE A   2      -5.724   3.347  -5.139  1.00  0.00           H
cATOM     23 HG21 ILE A   2      -3.455   0.048  -3.509  1.00  0.00           H
cATOM     24 HG22 ILE A   2      -3.677   1.472  -2.517  1.00  0.00           H
cATOM     25 HG23 ILE A   2      -5.044   0.479  -2.934  1.00  0.00           H
cATOM     26 HD11 ILE A   2      -7.022   1.890  -3.517  1.00  0.00           H
cATOM     27 HD12 ILE A   2      -5.896   2.370  -2.259  1.00  0.00           H
cccccccccccccccccccccccccccccccccccccccccccccccccccc

      go to 888

100   format(A12,A1,A1,A66)

333   continue

      close(2)
      close(1)

      stop
      end
```

**############### End of Pogram "remove_H_atoms.f" ###############**

**Program 15:**

**############### Start of Pogram "resd_num_reduct.f" ###############**

```
c     program name: ResidueNumberReduction.f

c c c c c c c c c c c c c c c c c c c c c c c c c c c c c c c c c c c c c c
c                                                                        c
```

```
c   Author:  Vicente M. Reyes, Ph.D., Asst. Professor                  c
c            Dept. of Biological Sciences, Sch. of Life Sciences       c
c            College of Sciences, Rochester Inst. of Technology        c
c            85 Lomb Memorial Dr., 08-1336                             c
c            Rochester, NY  14623-5603                                 c
c                                                                      c
c c c c c c c c c c c c c c c c c c c c c c c c c c c c c c c c c c c c c c

      character*2 resno
      character*2 nuresno
      character*22 left
      character*56 right

      open (unit =1, file = "filei")
      open (unit =2, file = "fileo")


888   read(1,100,end=333) left, resno, right
100   format(A22,A2,A56)


        if (resno.eq.'10') then
            nuresno = ' A'
        elseif (resno.eq.'11') then
            nuresno = ' B'
        elseif (resno.eq.'12') then
            nuresno = ' C'
        elseif (resno.eq.'13') then
            nuresno = ' D'
        elseif (resno.eq.'14') then
            nuresno = ' E'
        elseif (resno.eq.'15') then
            nuresno = ' F'
        elseif (resno.eq.'16') then
            nuresno = ' G'
        elseif (resno.eq.'17') then
            nuresno = ' H'
        elseif (resno.eq.'18') then
            nuresno = ' I'
        elseif (resno.eq.'19') then
            nuresno = ' J'
        elseif (resno.eq.'20') then
            nuresno = ' K'
        elseif (resno.eq.'21') then
            nuresno = ' L'
        elseif (resno.eq.'22') then
            nuresno = ' M'
        elseif (resno.eq.'23') then
            nuresno = ' N'
        elseif (resno.eq.'24') then
            nuresno = ' O'
        elseif (resno.eq.'25') then
            nuresno = ' P'
        elseif (resno.eq.'26') then
            nuresno = ' Q'
        elseif (resno.eq.'27') then
            nuresno = ' R'
        elseif (resno.eq.'28') then
            nuresno = ' S'
        elseif (resno.eq.'29') then
            nuresno = ' T'
        elseif (resno.eq.'30') then
            nuresno = ' U'
        elseif (resno.eq.'31') then
            nuresno = ' V'
        elseif (resno.eq.'32') then
```

```
        nuresno = ' W'
    elseif (resno.eq.'33') then
        nuresno = ' X'
    elseif (resno.eq.'34') then
        nuresno = ' Y'
    elseif (resno.eq.'35') then
        nuresno = ' Z'
    elseif (resno.eq.'36') then
        nuresno = ' a'
    elseif (resno.eq.'37') then
        nuresno = ' b'
    elseif (resno.eq.'38') then
        nuresno = ' c'
    elseif (resno.eq.'39') then
        nuresno = ' d'
    elseif (resno.eq.'40') then
        nuresno = ' e'
    elseif (resno.eq.'41') then
        nuresno = ' f'
    elseif (resno.eq.'42') then
        nuresno = ' g'
    elseif (resno.eq.'43') then
        nuresno = ' h'
    elseif (resno.eq.'44') then
        nuresno = ' i'
    elseif (resno.eq.'45') then
        nuresno = ' j'
    elseif (resno.eq.'46') then
        nuresno = ' k'
    elseif (resno.eq.'47') then
        nuresno = ' l'
    elseif (resno.eq.'48') then
        nuresno = ' m'
    elseif (resno.eq.'49') then
        nuresno = ' n'
    elseif (resno.eq.'50') then
        nuresno = ' o'
    elseif (resno.eq.'51') then
        nuresno = ' p'
    elseif (resno.eq.'52') then
        nuresno = ' q'
    elseif (resno.eq.'53') then
        nuresno = ' r'
    elseif (resno.eq.'54') then
        nuresno = ' s'
    elseif (resno.eq.'55') then
        nuresno = ' t'
    elseif (resno.eq.'56') then
        nuresno = ' u'
    elseif (resno.eq.'57') then
        nuresno = ' v'
    elseif (resno.eq.'58') then
        nuresno = ' w'
    elseif (resno.eq.'59') then
        nuresno = ' x'
    elseif (resno.eq.'60') then
        nuresno = ' y'
    elseif (resno.eq.'61') then
        nuresno = ' z'
    else
        nuresno = resno

  endif

  write(2,100) left, nuresno, right
```

```
      go to 888

333   continue

      close(2)
      close(1)

      stop

      end
```

**################   End of Pogram   "resd_num_reduct.f"   ###############**

**10. FIGURES:**

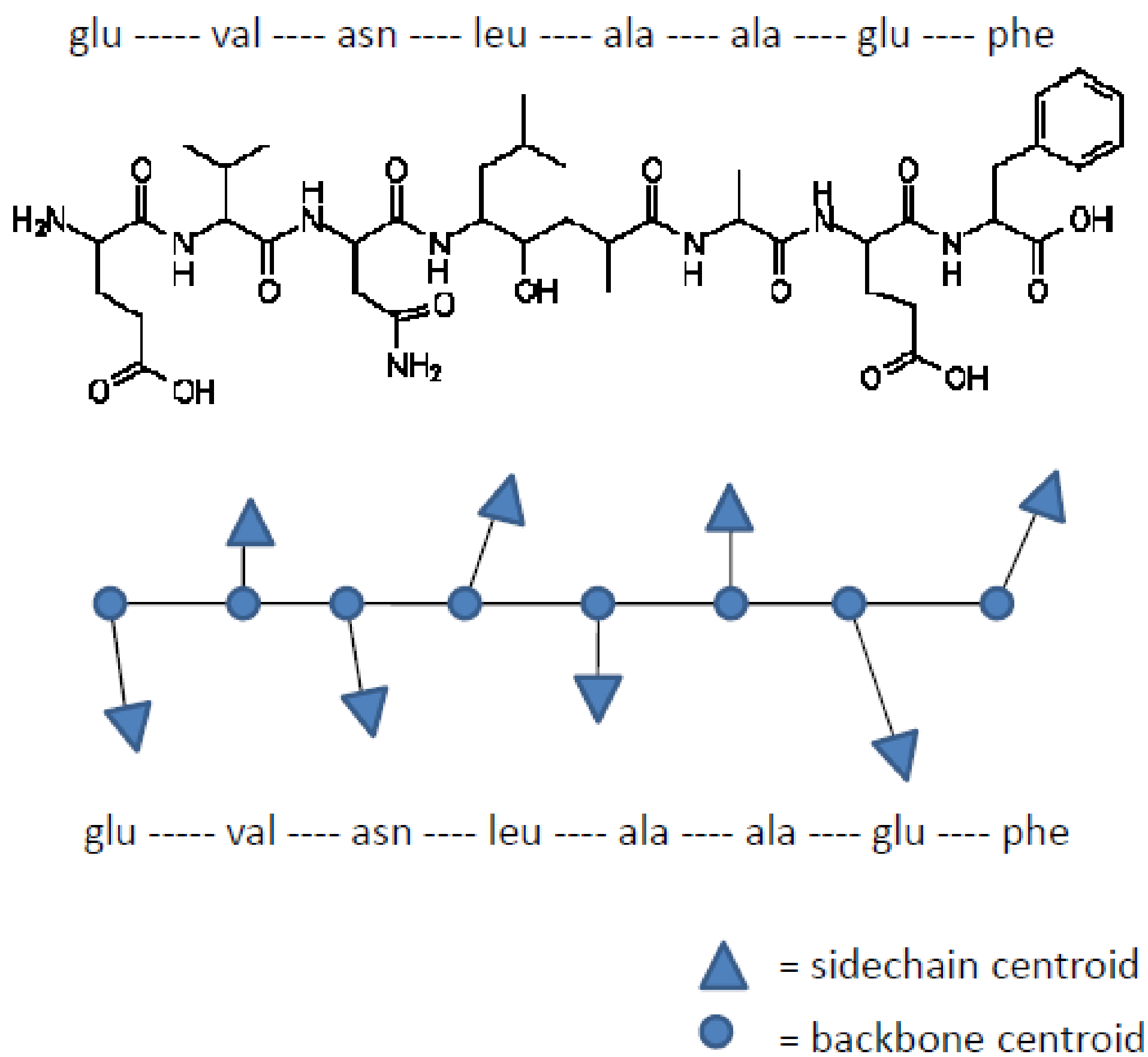

**Figure 1**

Figure 1 illustrates the use of the DCRR on the short peptide, OM99-2 an octapeptide memepsin inhibitor [from http://chemistry.umeche.maine.edu/CHY431/Peptidase17.html]. The amino acid sequence is shown above and below. The molecular structure in black above shows the peptide in all-atom representation; while the schematic representation in blue below is the double-centroid reduced representation. The blue triangle is the sidechain centroid of any particular amino acid in the peptide, while the blue square represents the centroid of the corresponding backbone atoms. Note that the atomicity of the peptide has been greatly reduced.

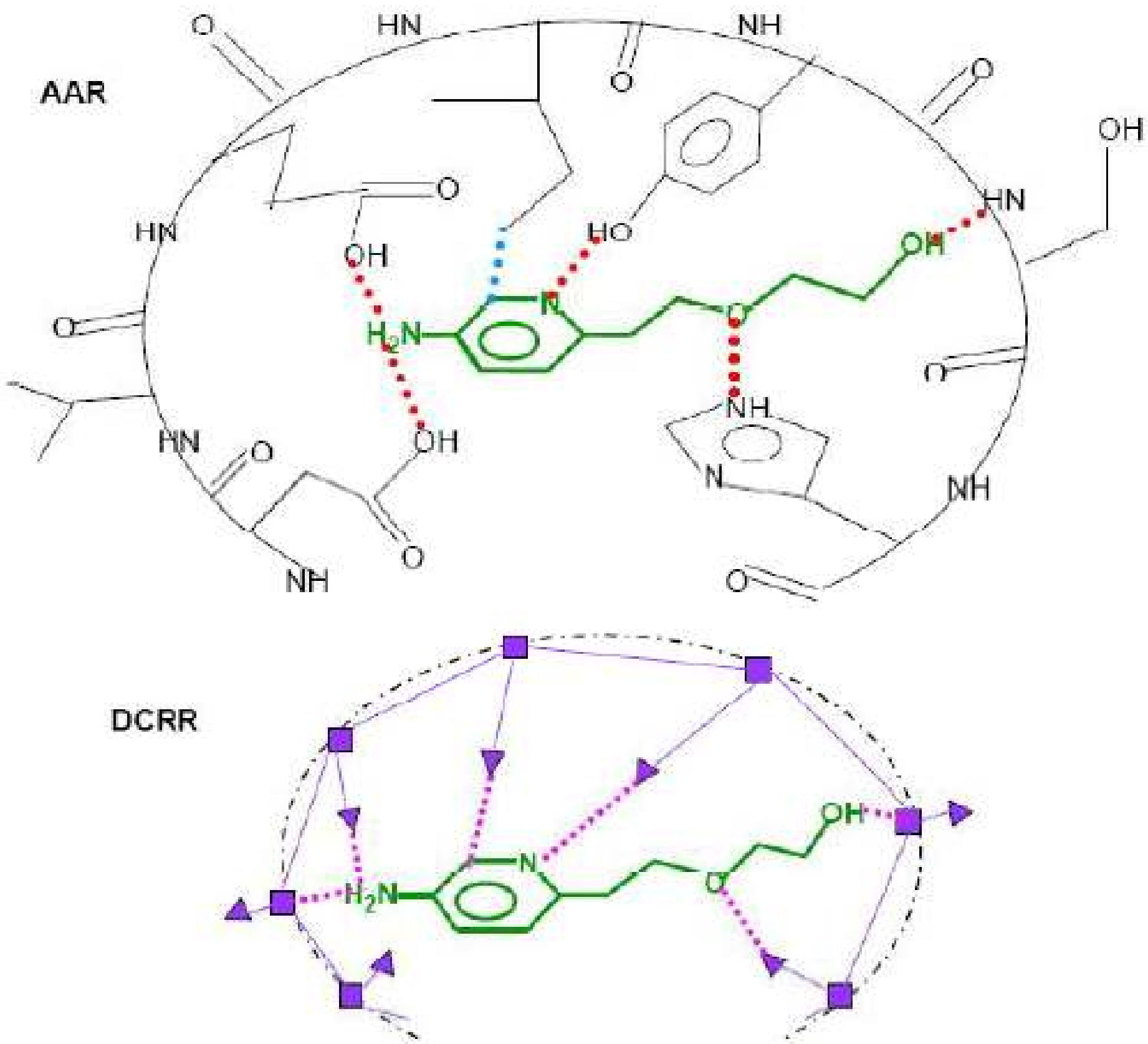

**Figure 2**

Figure 2 shows the ligand (in green) bound at the LBS; in the upper half, the protein is in AAR; while in the lower half, the protein is in DCRR (purple triangles as sidechain centroids and purple squares as backbone centroids). Broken lines represent hydrogen bonding or van der Waals interactions between protein and ligand. Note that when the protein is in AAR, the structure is more complicated and less amenable to modeling, while when it is in DCRR, the structure is much less complicated and more amenable to modeling.

**General 3D Search Motif Structure:**

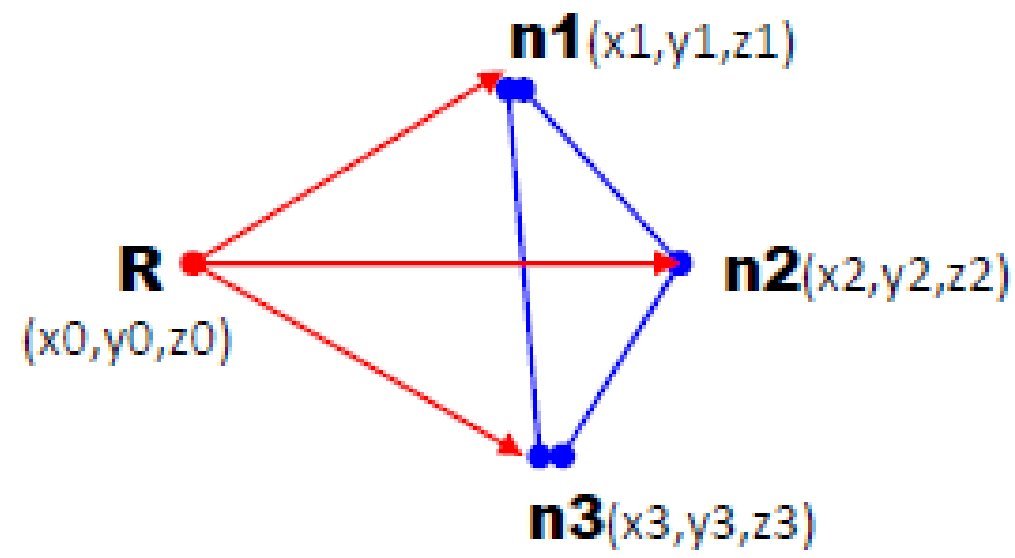

**Screening Algorithm Parameters:**

I. Qualitative parameters: at least 4; amino acid identities of:

  **R** == root

  **n1, n2, n3** == nodes

II. Quantitative parameters: at least 6; lengths of:

  **Rn1, Rn2, Rn3** == branches

  **n1n2, n2n3, n3n1** == node-edges

III. Fuzzy factor added to branch and node-edge lengths: **± ε** (~1.40 Å)

## Figure 3

Figure 3 shows the 3D tetrahedral search motif for a ligand. Root R and three nodes, n1, n2 and n3 are four amino acid in the binding site of the protein interacting with ligand atoms by hydrogen bonding or van der Waals interaction. Since the protein is in the DCRR, then the four vertices of the tetrahedron, R, n1, n2 and n3 must be sidechain or backbone centroids of amino acids in the LBS. Choice of root R and nodes n1, n2 and n3 are arbitrary but we usually choose R to be the most conserved or dominant amino acid in the LBS. In screening for LBSs, we add a "fuzzy factor" to the lengths of the sides Rn1, Rn2 Rn3, n1n2, n1n3 and n2n3 of the tetrahedron.

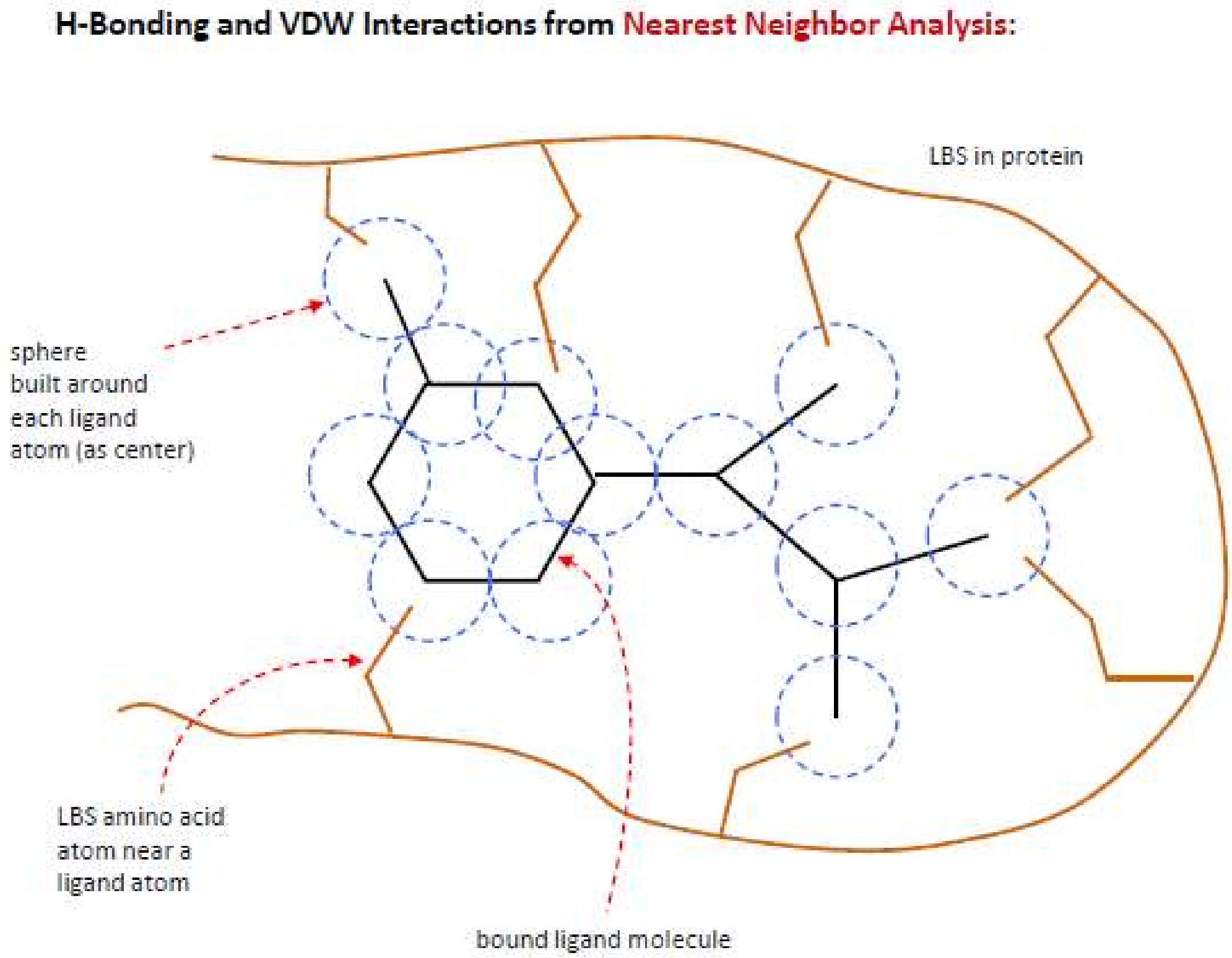

Figure 4

Figure 4 shows the concept behind nearest neighbor analysis. The ligand and the protein structure files (atomic coordinates) are separated into two files, then inputted into the nearest neighbor analysis program. The program iterates over each ligand atom, finding any protein atoms within a given radius of the ligand atom, typically set at 4.0Å (circles in blue broken lines). The results is a file containing the protein atom "neighbors" of each ligand atom. From this data, H-bonds and van der Waals interactions between ligand and protein atoms are determined by two other programs (find_Hbonds.f and find_VDW.f, respectively).

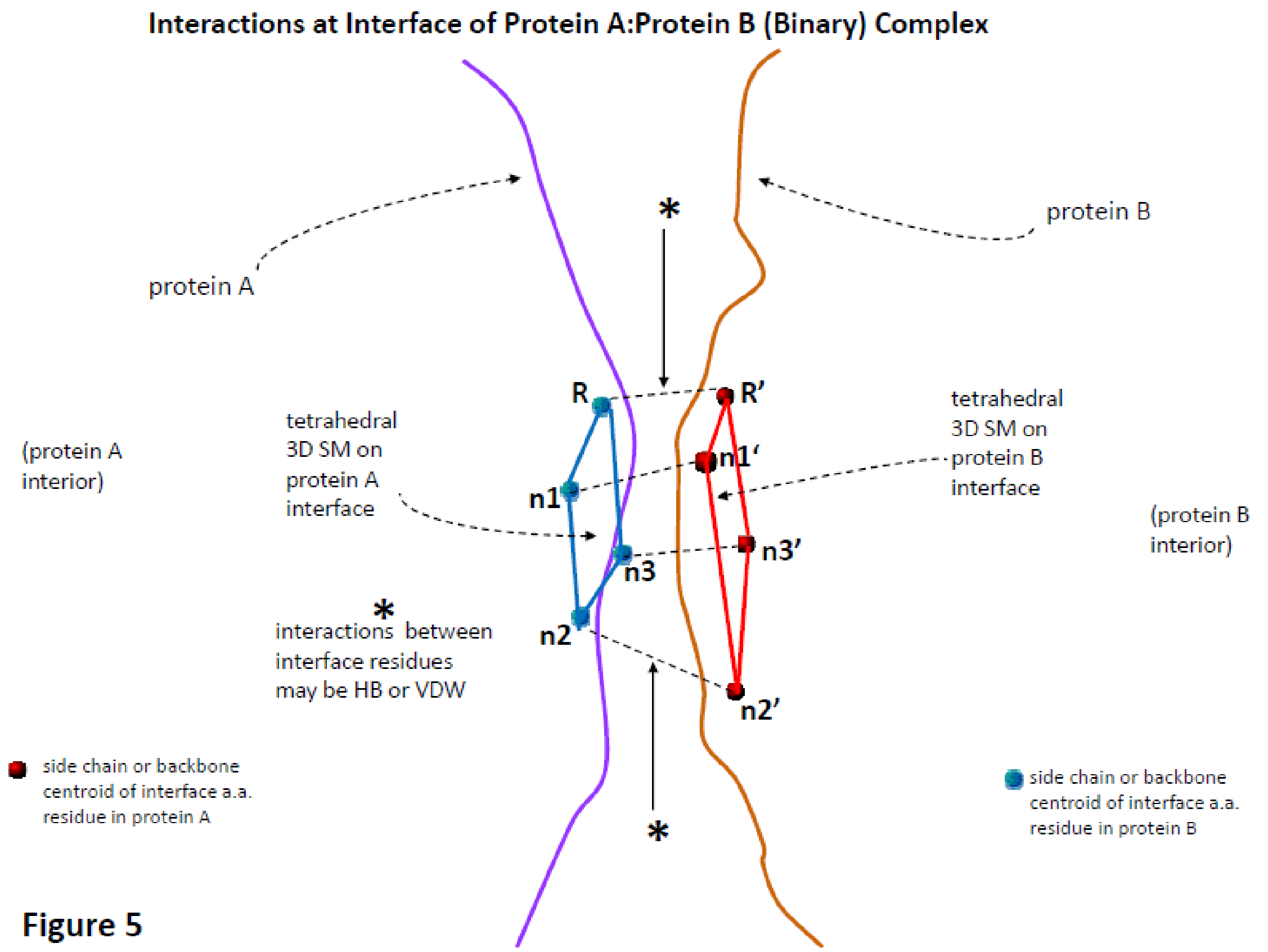

**Figure 5**

Figure 5 schematically shows the interface between two proteins A and B (purple and brown) engaged in PPI. This interface is specific. There are usually several hydrogen bonds and van der Waals interactions between the two proteins at their interface. The four most dominant is selected. The four atoms in protein A (blue dots on the protein A interface) and the corresponding four in protein B (red dots on the protein B interface) both form a tetrahedron. These two interacting tetrahedra on the PP interface is the 3D interface search motif tetrahedral pair (3D ISMTP) forms the basis for the screening of PPI partners in our procedure.

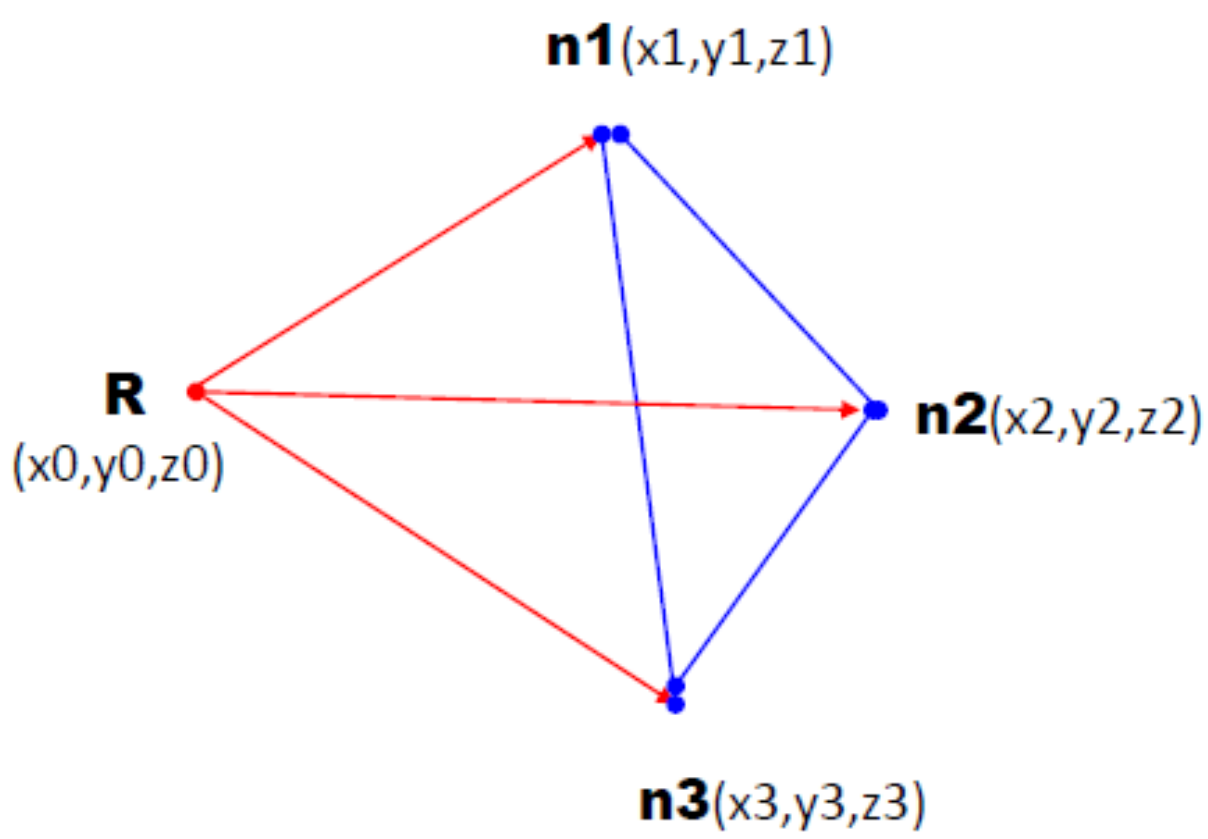

**3D Tetrahedral SM for:**

| | GTP (srGP) | Sialic Acid |
|---|---|---|
| R | D(s) | R(s)/Q(s) |
| n1 | A(b)/G(b) | Y(s)/S(s) |
| n2 | K(s) | W(s)/M(s) |
| n3 | G(b) | G(b)/V(b) |
| Rn1 | 6.975 | 5.949 |
| Rn2 | 6.951 | 9.590 |
| Rn3 | 10.732 | 7.823 |
| n1n2 | 8.544 | 5.396 |
| n1n3 | 8.002 | 6.600 |
| n2n3 | 5.650 | 5.206 |

**Figure 6**

Figure 6 shows the parameters for the screening of GTP (srGP) and sialic acid acid (SIA) for the 3D SM in our procedure. The root for the srGP 3D SM is the sidechain centroid of an asp residue, D(s), while that for SIA is the sidechain centroid of an arg or an asn residue ("/" indicates bijection, "or"), R(s)/Q(s). And similarly for nodes n1, n2 and n3. Branch Rn1 is 6.975 Å long in the srGP 3D SM, and 5.949 Å in the SIA 2D SM. And similarly for branches Rn2 and Rn3, and node-edges n1n2, n1n3 and n2n3.

**Complex D (cD) and Complex H (cH ) Interface 3D Search Motif Pairs**

motif components:

| | cD | cH |
|---|---|---|
| R = | N(b) | K(s) |
| n1 = | Q(s) | I(b) |
| n2 = | E(s) | E(s) |
| n3 = | G(b) | Q(s) |

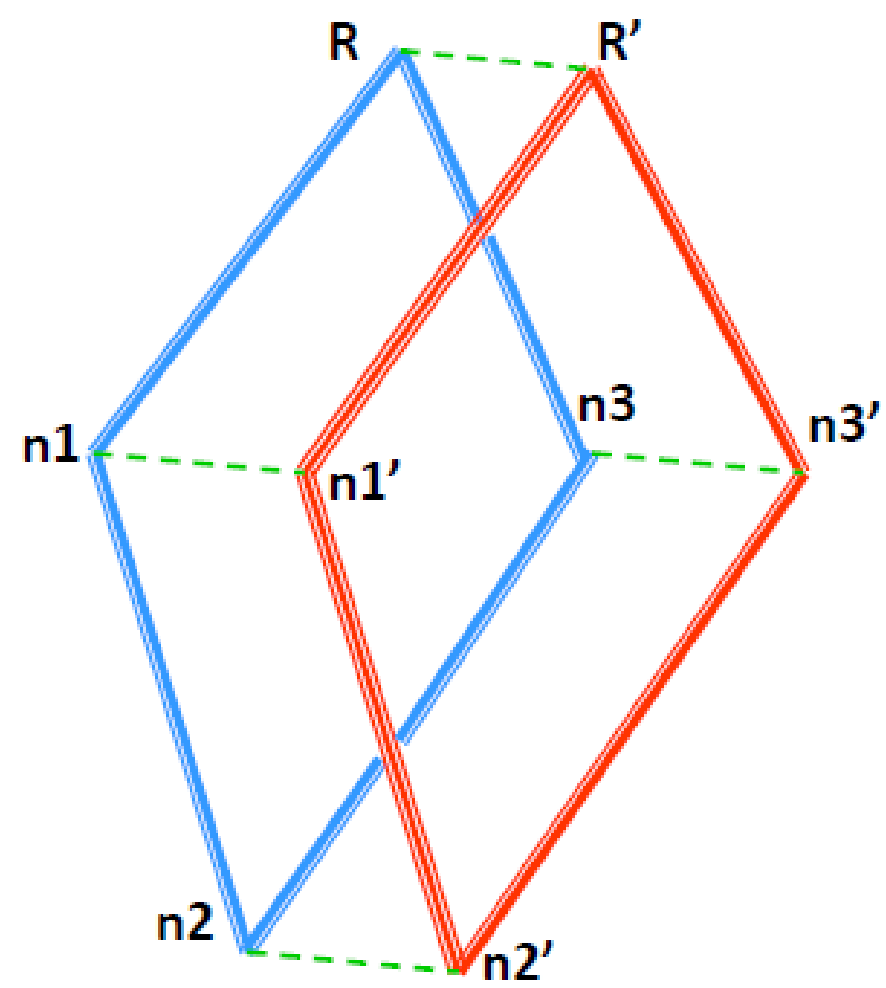

motif components:

| | cD | cH |
|---|---|---|
| R' = | R(s) | D(s) |
| n1' = | N(s) | E(s) |
| n2' = | S(s) | K(s) |
| n3' = | D(s) | D(b) |

| side | length, Å cD | length, Å cH |
|---|---|---|
| Rn1 | 5.464 | 5.602 |
| Rn2 | 15.631 | 10.855 |
| Rn3 | 12.442 | 11.814 |
| n1n2 | 18.923 | 10.129 |
| n1n3 | 14.339 | 8.854 |
| n2n3 | 6.038 | 18.362 |

| side | length, Å cD | length, Å cH |
|---|---|---|
| R'n1' | 6.453 | 8.306 |
| R'n2' | 15.374 | 7.371 |
| R'n3' | 11.068 | 16.069 |
| n1'n2' | 19.532 | 13.395 |
| n1'n3' | 15.996 | 8.303 |
| n2'n3' | 4.720 | 19.143 |

| side | length, Å cD | length, Å cH |
|---|---|---|
| RR' | 5.861 | 3.761 |
| n1n1' | 5.060 | 5.218 |
| n2n2' | 3.898 | 5.134 |
| n3n3' | 4.271 | 5.596 |

**Figure 7**

Figure 7 shows the two interacting tetrahedral from the interface of two proteins in binary complex with each other. Each is depicted as a square for ease of illustration, but in general they are indeed both tetrahedra. The blue one on the left (composed of nodes R, n1, n2, and n3) comes from one protomer and the red one on the right (composed of nodes R', n1', n2', and n3') comes from the other protomer. This figure refers to two binary complexes, complex D (cD, RAC complexed with P67PHOX RAC complexed with P67PHOX) and complex H (cH, Kinase-associated phosphatase (KAP) complexed w/phospho-CDK2). The lengths shown are in Angstroms, Å.

**11. TABLES:**

**All-Atom Representation (AAR) vs. Double-Centroid Reduced Representation (DCRR)**

AAR for leucine residue:

| | | | | | | | | | | | |
|---|---|---|---|---|---|---|---|---|---|---|---|
| ATOM | 398 | N | LEU | A | 908 | 35.964 | 49.482 | 10.054 | 1.00 | 45.16 | N |
| ATOM | 399 | CA | LEU | A | 908 | 35.004 | 48.670 | 9.322 | 1.00 | 44.49 | C |
| ATOM | 400 | C | LEU | A | 908 | 33.583 | 49.124 | 9.619 | 1.00 | 44.60 | C |
| ATOM | 401 | O | LEU | A | 908 | 32.722 | 49.122 | 8.741 | 1.00 | 44.40 | O |
| ATOM | 402 | CB | LEU | A | 908 | 35.180 | 47.188 | 9.671 | 1.00 | 43.60 | C |
| ATOM | 403 | CG | LEU | A | 908 | 36.535 | 46.572 | 9.291 | 1.00 | 43.23 | C |
| ATOM | 404 | CD1 | LEU | A | 908 | 36.551 | 45.108 | 9.692 | 1.00 | 43.18 | C |
| ATOM | 405 | CD2 | LEU | A | 908 | 36.782 | 46.711 | 7.791 | 1.00 | 43.00 | C |

DCRR for same leucine residue:

| | | | | | | | | | | | |
|---|---|---|---|---|---|---|---|---|---|---|---|
| ATOM | 398 | bbc | LEU | A | 908 | 34.318 | 49.100 | 9.434 | 1.00 | 45.16 | N |
| ATOM | 402 | sdc | LEU | A | 908 | 36.262 | 46.395 | 9.111 | 1.00 | 43.60 | C |

# Table 1.

Table 1, top panel, shows the a portion of a regular PDB file in all-atom representation (AAR) for a leucine residue. It has coordinates for all atoms (except H) in the residue: the four backbone atoms, N, CA, C' and O, as well as the sidechain atoms, CB. CG, CD1 and CD2, for a total of eight coordinates. When transformed into the double-centroid reduced representation (DCRR), this portion of the PDB becomes as shown on the lower panel: note that there are only two coordinates, one for the centroid of the backbone atoms, and another for that of the sidechain atoms, thus drastically reducing the atomicity (in this case by 75%).

# Table 2. Training Set for GTP-Binding Proteins (srGP):

| PDB ID | E.C. No. | Protein Description | Family |
|---|---|---|---|
| 1C4K | 4.1.1.17 | Ornithine Decarboxylase Mutant (Gly121Tyr) | N/A |
| 1E96 | N/A | Structure of RAC/P67Phos Complex | 02B |
| 1FRW | N/A | Structure of E. coli MOBA with Bound GTP and Mn | N/A |
| 1JFF | N/A | Refined Structure of Alpha-Beta Tubulin from Zinc-Induced Sheets Stabilized with Taxol | 02D |
| 1N6L | N/A | Crystal Structure of Human Rab5A A30P Mutant Complexed with GTP | 02B |
| 1NVU | N/A | Structural Evidence for Feedback Activation by Ras-GTP of the Ras-Specific Nucleotide Exchange Factor SOS | 02B |
| 1P16 | 2.7.7.50 | Structure of an mRNA Capping Enzyme Bound to the Phosphorylated Carboxyl Terminal Domain of RNA Polymerase II | 02G |
| 1TUB | N/A | Tubulin Alpha-Beta Dimer, Electron Diffraction | 02D |
| 1A9C | 3.5.4.16 | GTP-Cyclohydrolase I (C110S Mutant) in Complex w/ GTP | N/A |
| 1CKM | 2.7.7.50 | Structure of 2 Different Conformations of mRNA Capping Enzyme in Complex with GTP | 02G |
| 1HWX | 1.4.1.3 | Crystal Structure of Bovine Liver Glutamate Dehydrogenase Complexed w/ GTP, NADH & L-Glu | 02K |
| 1HWZ | 1.4.1.3 | Bovine Glutamate Dehydrogenase Complexed with NADPH, Glutamate and GTP | 02K |
| 1LOO | 6.3.4.4 | Crystal Structure of the Mouse Muscle Adenylo-succinate Synthetase Ligated with GTP | 02B |
| 1M7B | N/A | Crystal Structure of RND3/RHOE: Functional Implications | 02B |
| 1O3Y | N/A | Crystal Structure of Mouse ARF1(Delta17-Q71L), GTP Form | 02B |
| 2RAP | N/A | The Small G-Protein RAP2A in Complex with GTP | 02B |

Table 2 shows the identities of the training structures used for the screening for GTP- binding sites characteristic of the srGP family. There are 16 structures in all and each is of the small Ras-type G-protein family. The first column is the PDB ID of the structure deposited in the PDB; second column shows the E.C. (Enzyme Commission) code for the GTPAse ("N/A" if no GTPase activity); the third column gives a quick description of the protein; while the last column is the abbreviation used in the main paper where we described the full and complete study.

# Table 3. Training Set for Sialic Acid

## The Training Sets

Sialic Acid Binding Site:

| | | |
|---|---|---|
| 1JSN:A | *Influenza A virus* | Hemagglutinin HA1 chain (residues 1-325, chain A) and HA2 chain (residues 1-176, chain B) with bound N-acetyl-D-glucosamine (NAG), D-galactose (GAL), and O-sialic acid (SIA) |
| 1JSO:A | *Influenza A virus* | Hemagglutinin HA1 chain (residues 1-325; chain A) and HA2 chain (residues 1-176; chain B) with bound N-acetyl-D-glucosamine (NAG) and O-sialic acid (SIA) |
| 1W0O:A | *Vibrio cholerae* | Sialidase (E.C. 3.2.1.18; syn.: neuraminidase, nanase) with bound calcium ion, 2-deoxy-2,3-dehydro-N-acetyl-neuraminic acid (DAN) and O-sialic acid (SIA) |
| 1W0P:A | *Vibrio cholerae* | Sialidase (E.C. 3.2.1.18; syn.: neuraminidase, nanase) with bound calcium ion, glycerol (GOL), 2-amino-2-hydroxymethyl-propane-1,3-diol (TRS), and O-sialic acid (SIA) |
| 1MQN:A,D | *Influenza A virus* | Hemagglutinin HA1 chain (chains A, D, G) and HA2 chain (chains B, E, H) with bound N-acetyl-D-glucosamine (NAG), alpha-D-mannose (MAN), D-galactose (GAL) and O-sialic acid (SIA) molecules |

Table 3 shows the identities of the five training structures we used for the screening for SIA-binding sites, consisting of three hemagglutinin and two sialidase (neuraminidase) proteins. The first column gives the PDB ID of the structure, each of which has a bound sialic acid (SIA) molecule, and where the ":L" at the end (colon followed by a capital letter) indicates the chain ID(s) of the protein; the second column is the source organism; and the third column being the description of the protein in question.

**Table 4.**

**The Training Structures for Complexes D and H**

| | | | |
|---|---|---|---|
| **Complex D:** | **1E96:A,B** | **RAC complexed w/ P67PHOX** | |
| molecule 1 | *Homo sapiens* | Ras-related C3 botulinum toxin substrate 1 (syn.: RAC1) | |
| molecule 2 | *Homo sapiens* | neutrophil cytosol factor 2 (NCF-2) TPR domain, residues 1-203 (syn.: P67PHOX) | |
| **Complex H:** | **1FQ1:A,B** | **kinase-associated phosphatase (KAP) complexed w/ phospho-CDK2** | |
| molecule 1 | *Homo sapiens* | cyclin-dependent kinase inhibitor 3 (E.C.3.1.3.48) | |
| molecule 2 | *Homo sapiens* | cell division protein kinase 2 (E.C.2.7.1.-) | |

Table 4 shows the identities of the training structures used for the screening for protein-protein partners corresponding to RAC complexed with P67PHOX (complex D or cD) and kinase-associated phosphatase (KAP) complexed with phosphor-CDK2. Note that both binary complexes were from *Homo sapiens*. Both complexes have two protomers: for cD they are: Ras-related C3 botulinum substrate 1 a.k.a. RAC, and neutrophil factor (NF-2) TPR domain (residues 1-203), a.k.a. P67PHOX; while for cH, they are: cyclin-dependent kinase inhibitor3, and cell division protein kinase 2.

**Table 5.**

| PDB ID | LBS Detected | PDB Header | Source Organism | Remarks |
|---|---|---|---|---|
| 1RU8 | GTP | STRUCTURAL GENOMICS, UNKNOWN FUNCTION 11-DEC-03 | Pyrococcus furiosus | PUTATIVE N-TYPE ATP PYROPHOSPHATASE; NSG TARGET PFR23 |
| 1XTL | GTP | STRUCTURAL GENOMICS, UNKNOWN FUNCTION 22-OCT-04 | Bacillus subtilis | P104H MUTANT OF HYPOTHETICAL SUPEROXIDE DISMUTASE-LIKE PROTEIN YOJM |
| 1IUK | sialic acid | STRUCTURAL GENOMICS, UNKNOWN FUNCTION 05-MAR-02 | Thermus thermophilus | HYPOTHETICAL PROTEIN TT1466, A CONSERVED COA-BINDING PROTEIN |
| 1SQH | sialic acid | STRUCTURAL GENOMICS, UNKNOWN FUNCTION 18-MAR-04 | Drosophila malonogaster | HYPOTHETICAL PROTEIN CG14615-PA (Q9VR51), NSGC TARGET FR87 |
| 1VKA | sialic acid | STRUCTURAL GENOMICS, UNKNOWN FUNCTION 10-MAY-04 | Homo sapiens | HYPOTHETICAL PROTEIN Q15691: N-TERMINAL FRAGMENT MICROTUBULE-ASSOCIATED PROTEIN RP/EB FAMILY; SYN.: APC-BINDING PROTEIN EB1 |
| 1Y6Z | sialic acid | STRUCTURAL GENOMICS, UNKNOWN FUNCTION 07-DEC-04 | Plasmodium falciparum | C-TERMINAL DOMAIN OF PUTATIVE HEAT SHOCK PROTEIN PF14_0417 (CHAPERONE) |

Table 5 shows the results of our screening procedure for GTP and SIA binding sites. We selected 801 protein structures in the PDB in 2004 (see complete list in Reyes, V.M., 2015x) whose functions are not yet known (at that time). Using our procedure for LBS detection, we were able to pick out two and four proteins that are potentially GTP- (as srGP) and SIA binders, respectively, and these are shown above. Note that their PDB headers all say that their functions were yet unknown at the time (column 3). Note also that they come from diverse organisms (column 4).

**Table 6.**

| PDB ID | Interface 3D SM Detected | PDB Header | Source Organism | Remarks | Published Reference |
|---|---|---|---|---|---|
| 1J2R | complex D[f], monomer 1 | STRUCTURAL GENOMICS, UNKNOWN FUNCTION 09-JAN-03 | *Escherichia coli* | HYPOTHETICAL ISOCHORISMATASE FAMILY PROTEIN YECD WITH PARALLEL BETA-SHEET 3-2-1-4-5-6, ALPHA-BETA-ALPHA MOTIF | Nucleic Acids Res. 2004 Jul 1;32(Web Server issue):W606-9 |
| 2F4L | complex D, monomer 1 | STRUCTURAL GENOMICS, UNKNOWN FUNCTION 23-NOV-05 | *Thermotoga maritima* | PROTEIN TM0119, A PUTATIVE ACETAMIDASE | none |
| 1SBK | complex D, monomer 2 | STRUCTURAL GENOMICS, UNKNOWN FUNCTION 10-FEB-04 | *Escherichia coli* | HYPOTHETICAL PROTEIN YDII, NESGC TARGET ER29 | none |
| 1VH5 | complex D, monomer 2 | STRUCTURAL GENOMICS, UNKNOWN FUNCTION 01-DEC-03 | *Escherichia coli* | HYPOTHETICAL PROTEIN YDII, A PUTATIVE THIOESTERASE | none |
| 1VI8 | complex D, monomer 2 | STRUCTURAL GENOMICS, UNKNOWN FUNCTION 01-DEC-03 | *Escherichia coli* | HYPOTHETICAL PROTEIN YDII, A PUTATIVE THIOESTERASE | none |
| 1ZBR | complex D, monomer 2 | STRUCTURAL GENOMICS, UNKNOWN FUNCTION 08-APR-05 | *Porphyromonas gingivalis* | A CONSERVED HYPOTHETICAL ALPHA-BETA PROTEIN, PUTATIVE ARGININE DEIMINASE | none |
| 1XVS | complex H[g], monomer 1 | STRUCTURAL GENOMICS, UNKNOWN FUNCTION 28-OCT-04 | *Vibrio cholerae* | APAG PROTEIN | none |
| 1S7O | complex H, monomer 2 | STRUCTURAL GENOMICS, UNKNOWN FUNCTION 29-JAN-04 | *Streptococcus pygenes* | HYPOTHETICAL PROTEIN UPF0122, PUTATIVE DNA BINDING PROTEIN SP_1288; SRP, RNA POLYMERASE SIGMA FACTOR | Acta Crystallogr D Biol Crystallogr. 2004 Jul;60(Pt 7):1266-71. Epub 2004 Jun 22 |
| 1V99 | complex H, monomer 2 | STRUCTURAL GENOMICS, UNKNOWN FUNCTION 23-JAN-04 | *Pyrococcus horikoshii OT3* | PERIPLASMIC DIVALENT CATION TOLERANCE PROTEIN CUTA, WITH CUCL2 | none |
| 1XG8 | complex H, monomer 2 | STRUCTURAL GENOMICS, UNKNOWN FUNCTION 16-SEP-04 | *Staphylococcus aureus (APC23712)* | HYPOTHETICAL PROTEIN SA0789 | none |

[f] Complex D is RAC complexed w/ P67PHOX; RAC is monomer 1; P67PHOX is monomer 2

[g] Complex H is kinase-associated phosphatase (KAP) complexed w/ phospho-CDK2; KAP is monomer 1; phospho-CDK2 is monomer 2

Table 6 shows the results of our PPI partner prediction using our procedure, the interface search motif tetrahedral pair (3D ISMTP) method. Using again the 801 protein structures in the PDB whose functions were unknown, we screened the set twice: one for the first protomer of the complex, and a second time for the second protomer. We did this for both complexes D and H. The results are shown above. We detected two possible protomer 1 and four possible protomer 2 for complex D, for a total of 2x4 = 8 possible complex D's. We also detected one possible protomer 1 and three possible protomer 2 for complex H, for a total of 1x3 = 3 possible complex H.